\documentclass[aps,pra,11pt,tightenlines]{revtex4-1}
\usepackage{mathtools}
\usepackage{units}
\usepackage{amssymb}
\usepackage{amsmath}
\usepackage{amsthm}
\usepackage{graphicx} 
\usepackage{bm}
\usepackage{braket}
\usepackage[usenames,dvipsnames,svgnames,table]{xcolor}
\usepackage{marginnote}
\newcommand{\pr}{\nabla_{\vec{r}}}
\newcommand{\pR}{\nabla_{\vec{R}}}
\newcommand{\pt}{\partial_t}

\newcommand{\pop}{\vec{\hat{p}}}
\newcommand{\Pop}{\vec{\hat{P}}}
\newcommand{\popc}{\vec{\hat{p}}_{\rm c}}
\newcommand{\Popc}{\vec{\hat{P}}_{\rm c}}

\renewcommand{\vec}[1]{\bm{#1}}

\begin{document}
  
  \author{Axel Schild}
  \title{Electronic quantum trajectories with quantum nuclei}
  \affiliation{ETH Z\"urich,  Laboratorium f\"ur Physikalische Chemie,  8093 Z\"urich, Switzerland}
  
  \begin{abstract}
    Quantum trajectory calculations for electrons are a useful tool in the field of molecular dynamics, e.g.\ to understand processes in ultrafast spectroscopy.
    They have, however, two limitation:
    On the one hand, such calculations are typically based on the Born-Oppenheimer approximation (BOA) and the electron dynamics for stationary nuclei is considered, thus neglecting quantum effects of the nuclei.
    On the other hand, even if the quantum nuclear motion would be taken into account, a BOA dynamics on a single potential energy surface would not provide any electron trajectories because the electronic part is treated as a stationary problem.
    By using the exact factorization method, we overcome these limitations and generalize the theory of electronic quantum trajectories to a fully quantum-mechanical treatment of the nuclei.
    After reviewing the time-dependent theory of quantum hydrodynamics and quantum trajectories, we show that the nuclei can be viewed as a quantum clock for the electronic motion and we develop a fully quantum-mechanical clock-dependent version of quantum hydrodynamics.
    This theory is used to obtain electronic trajectories for quantum nuclei, as is exemplified for a model system of a proton-coupled electron transfer dynamics.
    Our work generalizes the concept of quantum trajectories and lays the foundations for the development of trajectory-based simulation methods of electron dynamics beyond the BOA.
  \end{abstract}
  
  \maketitle
  
  \section{Introduction}
  
  Trajectory-based methods are an important tool for molecules dynamics simulations.
  Such simulations are usually based on a Born-Oppenheimer picture \cite{born1927} to separate the nuclei from the electrons:
  The electronic problem is solved for given nuclear configurations, and the obtained energies of the electronic subsystem constitute potential energy surfaces for the nuclei.
  The nuclear problem can then be solved with trajectories either fully quantum-mechanically (e.g.\ with coupled coherent states\cite{shalashilin2004}), semi-classically (e.g.\ with surface hopping \cite{tully1990,subotnik2016}, multiple spawning \cite{ben-nun2000,lassmann2021}, or quantum-classical dynamics \cite{agostini2019,agostini2021}), or even classically (ab initio molecular dynamics \cite{iftimie2005}), as the nuclei are comparably heavy and quantum effects do not always play a relevant role.
  
  Although electrons are quantum particles, trajectory-based methods have been developed also for them.
  A typical application is ultrafast spectroscopy, where classical electron trajectories are used for high electron kinetic energies to compute observables of interest (see e.g.\ methods based on the strong-field approximation \cite{amini2019}, path integrals \cite{salieres2001}, or trajectory calculations for attosecond dynamics in liquids \cite{jordan2020,jelovina2021}).
  Additionally, quantum trajectories have been used as an analysis tool for the interpretation of electron dynamics \cite{takemoto2011,jooya2016,ivanov2017}.
  However, in all applications of electronic quantum trajectories known to the author, the quantum nature of the nuclei has not been considered.
  
  There is also a principal problem when one wants to calculate electronic quantum trajectories for moving nuclei within the Born-Oppenheimer picture.
  In the Born-Oppenheimer approximation (BOA), the nuclear dynamics is obtained from a single potential energy surface.
  To obtain this surface, the static electronic problem is solved and all information on the electron dynamics is lost \cite{barth2009,scherrer2013,schild2016}.
  This is a rather large drawback because many chemical reactions can described within the BOA but the lack of an electron dynamics prevents a detailed look on the reaction mechanism, except if symmetry constraints apply \cite{hege2010,andrae2011,hermann2016}.
  
  A trajectory-based simulation method for electrons and nuclei together can be computationally advantageous and has the potential to reveal information hidden in an electron dynamics with static nuclei or a nuclear dynamics with static electrons, e.g.\ regarding the reaction mechanism.
  In this article, we develop the theory of electronic quantum trajectories with a fully quantum-mechanical description of the nuclei.
  For this purpose, in Sec.\ \ref{sec:tdqhd} we briefly review the theory underlying quantum trajectories, i.e., quantum hydrodynamics \cite{madelung1926,madelung1927,renziehausen2018} and Bohmian mechanics \cite{broglie1927,bohm1952,bohm1952_2}.
  Those are derived from the time-dependent Schr\"odinger equation (TDSE), hence we call the theory time-dependent quantum hydrodynamics (TDQHD).
  In Sec.\ \ref{sec:cdse} we then introduce the exact factorization \cite{abedi2010,abedi2012}.
  It separates the molecular wavefunction into a marginal nuclear part and a conditional electronic part, very similar to what is done in the BOA, but without making an approximation.
  The determining equation of the electronic wavefunction in the exact factorization is a generalized Schr\"odinger equation where the nuclei play the role of a quantum clock instead of a classical external time parameter.
  We thus call this equation the clock-dependent Schr\"odinger equation (CDSE) \cite{schild2018}.
  In Sec.\ \ref{sec:cdqhd}, clock-dependent quantum hydrodynamics (CDQHD) is derived from the CDSE and quantum trajectories for the electrons are obtained for quantum nuclei.
  Those trajectories are illustrated in Sec.\ \ref{sec:example} with a model system for proton-coupled electron transfer.
  Sec.\ \ref{sec:discussion} provides a discussion about the relevance of electronic quantum trajectories and gives ideas for future research directions.
  
%   vanishing EFD
%   
%   For practical applications, Bohmian mechanics can be viewed as trajectory-based quantum hydrodynamics, and in the last years a number of  studies have appeared that aim at using these quantum trajectories to develop simulation methods, in particular for molecular dynamics \cite{wyatt2001,lopreore2002,wyatt2005,rassolov2005,curchod2013,agostini2018}, also by extending it to complex-valued trajectories \cite{goldfarb2006,zamstein2012,zamstein2012_2,koch2017} or by using the ``conditional wavefunction'' approach \cite{oriols2007,benseny2014,albareda2014,albareda2016}.
%   
%   the definition of a tunneling time \cite{steinberg1995,landsman2015,zimmermann2016,sokolovski2018}
  
  \section{Time-dependent quantum hydrodynamics}
  \label{sec:tdqhd}
  
  In this section we briefly present TDQHD \cite{madelung1926,madelung1927} and the concept of quantum trajectories \cite{broglie1927,bohm1952,bohm1952_2,wyatt2005}.
  Our aim is to review the main ideas and to introduce the notation for the discussion of CDQHD.
  
  We start from the  single-particle TDSE
  \begin{align}
    i \hbar \pt \varphi(\vec{r}|t) = \left(\frac{\pop^2}{2 m} + V(\vec{r},t) \right) \varphi(\vec{r}|t)
    \label{eq:tdse}
  \end{align}
  where the wavefunction $\varphi \in \mathbb{C}$ descibes the state of the particle.
  Here, $\pop = -i \hbar \pr$ is the momentum operator, $m$ is the mass of the particle, $t \in \mathbb{R}$ is the time parameter, $\pt$ is the derivative with respect to (w.r.t.) $t$, $\vec{r} \in \mathbb{R}^3$ are the coordinates of the particle and $\pr$ is the gradient w.r.t.\ $\vec{r}$.
  The scalar potential $V \in \mathbb{R}$ represents the interaction with an unspecified environment.
  
  To obtain TDQHD we use the polar form 
  \begin{align}
    \varphi(\vec{r}|t) 
      &= \sqrt{\rho(\vec{r}|t)} e^{i s(\vec{r}|t)}        \label{eq:varphi_polar}  \\
      &= e^{w(\vec{r}|t) + i s(\vec{r}|t)} \label{eq:varphi_polar2}
  \end{align}
  of the wavefunction with the probability density $\rho = |\varphi|^2 \in \mathbb{R}$ , the classical action $s \in \mathbb{R}$ (it is dimensionless here but usually given in units of $\hbar$), and the quantum action $w \in \mathbb{R}$.
  We call an object ``quantum'' if it vanishes when a suitable classical limit (see \cite{klein2012,eich2016}) is taken for $\varphi$ and ``classical'' otherwise.
  After inserting \eqref{eq:varphi_polar} into the TDSE \eqref{eq:tdse} and separating the result into real and imaginary parts, two equations are obtained:
  The first equation is a Hamilton-Jacobi equation
  \begin{align}
    0 = \hbar \pt s(\vec{r}|t) + H(\vec{p},\vec{r}|t)
    \label{eq:tdhje}
  \end{align}
  that defines particle trajectories.
  The Hamiltonian function appearing in \eqref{eq:tdhje} is
  \begin{align}
    H(\vec{p},\vec{r}|t) &= \frac{\vec{p}(\vec{r}|t)^2}{2 m} + V(\vec{r}|t) + u(\vec{r}|t)
  \end{align}
  and contains the classical momentum field
  \begin{align}
    \vec{p}(\vec{r}|t) = \hbar \pr s(\vec{r}|t) \in \mathbb{R}
    \label{eq:p}
  \end{align}
  as well as the quantum potential
  \begin{align}
    u(\vec{r}|t) 
     = -\frac{\hbar^2}{2 m} \frac{\pr^2 |\varphi|}{|\varphi|}.
    \label{eq:qpot0}
  \end{align}
  The second equation is a continuity equation 
  \begin{align}
    0 = \pt \rho(\vec{r}|t) + \pr \cdot \vec{j}(\vec{r}|t)
    \label{eq:tdce}
  \end{align}
  that represents the conservation of the probability density $\rho$ \cite{aris1989}.
  It is equivalent to the statement that the change of density $\pt \rho$ in a volume $\Omega$ is given by the flow of density through the boundary of $\Omega$.
  That flow is given by the probability flux (current) density 
  \begin{align}
    \vec{j}(\vec{r}|t) = \rho(\vec{r}|t) \frac{\vec{p}(\vec{r}|t)}{m}.
    \label{eq:fd}
  \end{align}
  
  To compare with the clock-dependent case discussed below, we express \eqref{eq:tdhje} and \eqref{eq:tdce} in terms of momentum fields alone.
  For this purpose, we introduce the classical momentum operator 
  \begin{align}
    \popc &= i \pop = \hbar \pr
    \label{eq:popc}
  \end{align}
  and the quantum momentum field (see also \cite{garashchuk2014,gao2017})
  \begin{align}
    \vec{\pi}(\vec{r}|t) = \frac{\hbar \pr \rho}{2 \rho} = \hbar \pr w(\vec{r}|t)
  \end{align}
  that represents the relative variation $\frac{\pr \rho}{\rho}$ of the density with coordinate $\vec{r}$. 
  With these definitions, we write the quantum potential \eqref{eq:qpot0} as 
  \begin{align}
    u(\vec{r}|t) 
     = -\frac{1}{2 m} \left( \popc \cdot \vec{\pi}(\vec{r}|t) + \vec{\pi}(\vec{r}|t)^2 \right).
    \label{eq:qpot}
  \end{align}
  This form shows explicitly that the quantum potential is an additional kinetic term originating from the variation of $\rho$, i.e., from the confinement of the quantum particle.
  We also write the continuity equation \eqref{eq:tdce} as
  \begin{align}
    0 
%       &= \hbar^2 \left( \pt w + \frac{\pr \cdot \vec{j}}{2 \rho} \right)
      &=\hbar^2 \pt w + \frac{1}{m} \left( \vec{\pi} \cdot \vec{p} + \frac{\popc \cdot \vec{p}}{2} \right).
    \label{eq:tdce1}
  \end{align}
  While version \eqref{eq:tdce} of the continuity equation contains quantities that are non-zero only in regions where the particle can be found, version \eqref{eq:tdce1} is obtained from \eqref{eq:tdce} after division by $\rho$ and, thus, relates quantities that can be sizable everywhere.
  
  To obtain quantum trajectories, the Hamilton-Jacobi equation \eqref{eq:tdhje} is solved with the method of characteristics \cite{agostini2018}.
  By using this method, \eqref{eq:tdhje} is interpreted as differential equation for the phase $s$ alone, i.e., it is assumed that the quantum potential $u$ (or $\rho$ or $w$) is known.
  Then, \eqref{eq:tdhje} is solved along parametrized curves.
  The parameter can be identified with $t$ and the curves can be obtained by solving
  \begin{align}
    \frac{d}{dt} \vec{r}_{\rm t}(t) &= \frac{\vec{p}_{\rm t}(t)}{m} 
      \label{eq:traj_x} \\
    \frac{d}{dt}  \vec{p}_{\rm t}(t) &= -\pr H(\vec{p}_{\rm t}(t),\vec{r}_{\rm t}(t)|t) 
      \label{eq:traj_p}\\
    \frac{d}{dt}  S_{\rm t}(t)       &= \frac{\left( \vec{p}_{\rm t}(t) \right)^2}{m} - H(\vec{p}_{\rm t}(t),\vec{r}_{\rm t}(t)|t).
      \label{eq:traj_S}
  \end{align}
  Here, $\vec{r}_{\rm t}$, $\vec{p}_{\rm t}$, and $S_{\rm t}$ are the values of $\vec{r}$, $\vec{p}$, and $S$ along the trajectory, respectively.
  From \eqref{eq:traj_x} it is possible to interpret the curves $\vec{r}_{\rm t}(t)$ as trajectories of the particle that are guided by the wavefunction $\varphi$ via the momentum field $\vec{p}$ or, if the momentum field is computed from \eqref{eq:traj_p}, via the quantum potential $u$ appearing in the Hamiltonian function $H$.
  If several trajectories are considered with their initial locations $\vec{r}_{\rm t}(t_0)$ randomly sampled according to $\rho(\vec{r}|t_0)$ at some time $t_0$ (or if the trajectories are equally spaced but carry a weight according to this distribution) the continuity equation \eqref{eq:tdce} ensures that the trajectories yield the distribution $\rho(\vec{r}|t)$ for any time $t$.
  Hence, the basic equations of quantum hydrodynamics give rise to a particle picture of quantum mechanics where the particles have a definite trajectory in space, but the trajectories are guided by the phase of the wavefunction (or the quantum potential) and distributed according to its squared magnitude.
  
  \section{The exact factorization}
  \label{sec:cdse}
  
  In the exact factorization \cite{hunter1975,abedi2010,abedi2012}, we consider a wavefunction $\psi(\vec{R},\vec{r})$ depending on two sets of particle coordinates.
  Below, $\vec{R}$ are the nuclei and $\vec{r}$ are the electrons, but there is no principle restriction on the type and number of particles that these coordinates shall represent. 
  To simplify the equations, however, we restrict the presentation here to two particles with masses $M$ and $m$.
  The Schr\"odinger equation for these particles is
  \begin{align}
    \left( \frac{\Pop^2}{2M} + \frac{\pop^2}{2m} + V(\vec{R},\vec{r}) \right) \psi(\vec{R},\vec{r}) = E \psi(\vec{R},\vec{r}),
    \label{eq:tise_psi}
  \end{align}
  where $V(\vec{R},\vec{r})$ is a scalar potential and $\Pop = -i \hbar \pR$ and $\pop = -i \hbar \pr$ are two momentum operators (we could also add vector potentials to $\Pop$ and $\pop$ without changing the results discussed below).
  Next, the joint probability density $|\psi|^2$ is written as product
  \begin{align}
    |\psi(\vec{R},\vec{r})|^2 = |\chi(\vec{R})|^2 |\phi(\vec{r}|\vec{R})|^2,
  \end{align}
  where 
  \begin{align}
    |\chi(\vec{R})|^2 := \Braket{\psi(\vec{R},\vec{r})|\psi(\vec{R},\vec{r})}
    \label{eq:chi}
  \end{align}
  is the marginal probability density of finding the particle of mass $M$ at $\vec{R}$ independent of where the particle of mass $m$ is.
  The symbol $\Braket{\cdot}$ indicates the scalar product w.r.t.\ the coordinates $\vec{r}$.
  The function $\chi$ is called marginal wavefunction.
  The probability density $|\phi(\vec{r}|\vec{R})|^2$ is the conditional probability density for finding the particle with mass $m$ at $\vec{r}$ given the particle with mass $M$ is at $\vec{R}$.
  The conditional wavefunction is defined as 
  \begin{align}
    \phi(\vec{r}|\vec{R}) := \frac{\psi(\vec{R},\vec{r})}{\chi(\vec{R})}
    \label{eq:phidef}
  \end{align}
  and obeys the partial normalization condition
  \begin{align}
    \langle \phi(\vec{r}|\vec{R}) | \phi(\vec{r}|\vec{R}) \rangle \stackrel{!}{=} 1
  \end{align}
  for all $\vec{R}$.
  The equations of motion for $\chi$ and $\phi$ that follow are
  \begin{align}
    E       \chi &= \left( \frac{(\Pop + \vec{A}(\vec{R}))^2}{2M} + \epsilon(\vec{R}) \right) \chi \\
    \hat{C} \phi &= \left( \frac{\pop^2}{2m} + V(\vec{R},\vec{r}) + \hat{U} - \epsilon(\vec{R}) \right) \phi \label{eq:phi}
  \end{align}
  with scalar and vector potentials
  \begin{align}
    \epsilon(\vec{R}) &:= \Braket{\phi|\frac{\pop^2}{2m} + V(\vec{R},\vec{r}) + \hat{U} - \hat{C}|\phi} \\
    \vec{A}           &:= \Braket{\phi|\Pop \phi},
  \end{align}
  respectively, with the kinetic operator
  \begin{align}
    \hat{U} &= \frac{(\Pop-\vec{A})^2}{2M}
  \end{align}
  and with the coupling operator
  \begin{align}
    \hat{C} &= -\frac{1}{M} \frac{(\Pop+\vec{A}) \chi(\vec{R})}{\chi(\vec{R})} \cdot \left( \Pop - \vec{A} \right)
  \end{align}
  that depends explicitly on the wavefunction $\chi$.
  The choice of the phase of $\chi$ is a gauge freedom, i.e., the transformation 
  \begin{align}
    \chi'(\vec{R})         &= \chi(\vec{R}) e^{-i \theta(\vec{R})}        \\
    \phi'(\vec{r}|\vec{R}) &= \phi(\vec{r}|\vec{R}) e^{i \theta(\vec{R})} \\
    \vec{A}'(\vec{R})      &= \vec{A}(\vec{R}) + \pR \theta(\vec{R})
  \end{align}
  leaves the total wavefunction $\psi$ as well as all 
  equations of motion in the theory invariant.
  
  In the following, we use an important insight:
  The time parameter that appears in quantum mechanics is a classical conditional parameter, and time-dependent quantum mechanics can be derived from time-independent quantum mechanics via the exact factorization and/or an approximation similar to the BOA \cite{briggs2000,briggs2015,schild2018}.
  To show that, the static system described by $\psi(\vec{R},\vec{r})$ is partitioned into a clock with wavefunction $\chi(\vec{R})$ and a system with wavefunction $\phi(\vec{r}|\vec{R})$ that conditionally depends on the configuration $\vec{R}$ of the clock.
  If the clock behaves classically, the equation of motion for $\chi$ becomes the analogue of a time-independent Hamilton-Jacobi equation \cite{briggs2000,briggs2015,eich2016,schild2018}.
  Solving this equation for its characteristics yields classical trajectories.
  The classical configuration and momentum along these trajectories define a parameter $t$, the time.
  In the equation of motion for $\phi$, \eqref{eq:phi}, the operator $\hat{U}$ vanishes, the coupling operator becomes
  \begin{align}
    \hat{C} \rightarrow i \hbar \vec{V}_{\rm cl} \nabla_{\vec{R}_{\rm cl}} \rightarrow i \hbar \pt
  \end{align}
  with classical configuration $\vec{R}_{\rm cl}(t)$ and velocity $\vec{V}_{\rm cl}(t) = \pt \vec{R}_{\rm cl}(t)$ of the particle at $\vec{R}$, the potential $\epsilon(\vec{R}_{\rm cl}(t))$ becomes a purely $t$-dependent function that can be removed by a phase transformation, and \eqref{eq:phi} turns into the TDSE 
  \begin{align}
    i \hbar \pt \phi(\vec{r}|t) = \left( \frac{\pop^2}{2m} + V(\vec{R},\vec{r}) \right) \phi(\vec{r}|t)
  \end{align}
  for the particle at $\vec{r}$ with wavefunction $\phi(\vec{r}|\vec{R}_{\rm cl}(t)) \rightarrow \phi(\vec{r}|t)$.
  
  Consequently, in the classical limit \cite{klein2012,eich2016} for $\chi(\vec{R})$ the particle at $\vec{R}$ acts as a classical clock on which the the state $\phi(\vec{r}|\vec{R})$ conditionally depends.
  If the classical limit is not taken, the marginal wavefunction $\chi$ is the wavefunction of a quantum clock for the system described by $\phi$.
  Consequently, the conditional equation \eqref{eq:phi} is the quantum-mechanical analogue of the semi-classical TDSE where a quantum clock is considered instead of a classical time.
  We thus call \eqref{eq:phi} the CDSE \cite{schild2018}.
  The particle at $\vec{r}$ is the (quantum) system and the particle at $\vec{R}$ is the (quantum) clock in the following.
  
  \section{Clock-dependent quantum hydrodynamics}
  \label{sec:cdqhd}

  The basic equations of time-dependent quantum hydrodynamics, i.e., the Hamilton-Jacobi equation \eqref{eq:tdhje} and the continuity equation \eqref{eq:tdce}, are derived from the TDSE. 
  Similar clock-dependent versions of these equations can be found by starting from the CDSE. 
  The derivation proceeds along the same lines as for the time-dependent case, i.e., the conditional wavefunction $\phi$ is written in its polar form and inserted into the CDSE, the equation is separated into its 
  real and imaginary parts, and the resulting equations are expressed in terms of the probability density $|\phi|^2$, momentum densities, and probability flux densities.
  Before stating the results of this procedure, a few quantities need to be defined.
  
  For a generic function $\varphi(\vec{R},\vec{r}) = e^{w(\vec{R},\vec{r}) + i s(\vec{R},\vec{r})} \in \mathbb{C}$ with $s, w \in \mathbb{R}$, two momentum fields w.r.t.\ $\vec{R}$ are defined (using uppercase letters).
  The first is the classical momentum field
  \begin{align}
    \vec{P}[\varphi,\vec{A}] 
      :=\hbar \frac{\operatorname{Im} (\bar{\varphi} \pR \varphi)}{|\varphi|^2} + \vec{A}
       = \hbar \pR s + \vec{A}
  \end{align}
  and the second is the quantum momentum field
  \begin{align}
    \vec{\Pi}[\varphi] 
      := \hbar \frac{\operatorname{Re} (\bar{\varphi} \pR \varphi)}{|\varphi|^2}
      = \hbar \pR w.
  \end{align}
%   Both fields are real-valued, $\{\vec{P}, \vec{\Pi}\} \in \mathbb{R}$.
  The flux density corresponding to $\vec{P}$ is
  \begin{align}
    \vec{J}[\varphi,\vec{A}] &= \frac{1}{M} |\varphi|^2 \vec{P}[\varphi,\vec{A}].
  \end{align}
  Similar fields are defined w.r.t.\ $\vec{r}$ (using lowercase letters), i.e., the classical momentum field
  \begin{align}
    \vec{p}[\varphi] 
      := \hbar \frac{\operatorname{Im} (\bar{\varphi} \pr \varphi)}{|\varphi|^2}
       = \hbar \pr s
  \end{align}
  and the quantum momentum field
  \begin{align}
    \vec{\pi}[\varphi] 
      := \hbar \frac{\operatorname{Re} (\bar{\varphi} \pr \varphi)}{|\varphi|^2}
      = \hbar \pr w.
  \end{align}
%   with $\{ \vec{p}, \vec{\pi} \} \in \mathbb{R}$.
  The corresponding flux density is 
  \begin{align}
    \vec{j}[\varphi] &= \frac{1}{m} |\varphi|^2 \vec{p}[\varphi].
  \end{align}
  Finally, we define the probability density for the system depending on 
  $\vec{r}$ and conditionally on $\vec{R}$ as
  \begin{align}
    \rho(\vec{r}|\vec{R}) = |\phi(\vec{r}|\vec{R})|^2.
  \end{align}
  
  With all these definitions, we can now give the equations of CDQHD.
  The first is the clock-dependent Hamilton-Jacobi equation (CDHJE)
  \begin{align}
    0 = \frac{1}{M} 
        \left( \vec{P}[\chi,\vec{A}] \cdot \vec{P}[\phi,-\vec{A}] 
             - \vec{\Pi}[\chi] \cdot \vec{\Pi}[\phi] \right) 
    - \epsilon(\vec{R})\nonumber 
       + H_{\rm S }(\vec{r}|\vec{R}) 
        + H_{\rm SC}(\vec{r}|\vec{R}) %\\
    \label{eq:cdhje}
  \end{align}
  with the Hamiltonian function of the system
  \begin{align}
    H_{\rm S }(\vec{r}|\vec{R}) 
      &:= \frac{\left(\vec{p}[\phi]\right)^2}{2 m} + u(\vec{R},\vec{r}) + V(\vec{R},\vec{r})
  \end{align}
  and the system(clock) Hamiltonian function
  \begin{align}
    H_{\rm SC}(\vec{r}|\vec{R})
      &:= \frac{\left(\vec{P}[\phi,-\vec{A}]\right)^2}{2 M} + U(\vec{R},\vec{r}).
  \end{align}
  Here, system(clock) means that the quantity describes the system (the particle(s) at $\vec{r}$ with wavefunction $\phi(\vec{r}|\vec{R})$) when there is a change in the coordinates $\vec{R}$ of the clock, i.e., derivatives w.r.t.\ $\vec{R}$ are taken.
  The Hamiltonian functions contain the quantum potentials of the system
  \begin{align}
    u(\vec{R},\vec{r}) = -\frac{1}{2m} \left( \popc \cdot \vec{\pi}[\phi] + \left(\vec{\pi}[\phi]\right)^2 \right),
  \end{align}
  and the system(clock) quantum potential
  \begin{align}
    U(\vec{R},\vec{r}) = -\frac{1}{2M} \left( \Popc \cdot \vec{\Pi}[\phi] + \left(\vec{\Pi}[\phi]\right)^2 \right).
  \end{align}
  The classical momentum operators used in these equations are
  \begin{align}
    \popc &= i \pop = \hbar \pr \\
    \Popc &= i \Pop = \hbar \pR
  \end{align}
  
  Although \eqref{eq:cdhje} looks rather different than its time-dependent version \eqref{eq:tdhje}, the equations are similar.
  In particular, there are the correspondences
  \begin{align}
    \frac{1}{M} \vec{P}[\chi,\vec{A}] \cdot \vec{P}[\phi,-\vec{A}] \leftrightarrow \hbar \pt S(\vec{r}|t) \label{eq:cor1} \\
    H_{\rm S }(\vec{p}[\phi],\vec{r}|\vec{R})          \leftrightarrow H(\vec{p},\vec{r}|t).
  \end{align}
  For $\vec{A} = \vec{0}$, the first correspondence becomes simply
  \begin{align}
    \frac{1}{M} \vec{P}[\chi] \cdot \pR \leftrightarrow \pt.
  \end{align}
  Some additional terms also appear in the fully quantum-mechanical equation \eqref{eq:cdhje} that have no equivalent in the time-dependent Hamilton-Jacobi equation \eqref{eq:tdhje}:
  Those are $\vec{\Pi}[\chi] \cdot \vec{\Pi}[\phi]$ that connects the clock and the system(clock) quantum momentum fields as well as the system(clock) Hamiltonian function $H_{\rm SC}$.
  
  The second equation of CDQHD is the clock-dependent continuity equation (CDCE),
  \begin{align}
    0 =& \frac{1}{M} \vec{P}[\chi,\vec{A}] \cdot \pR \rho
       + \pr \cdot \vec{j}[\phi] \nonumber \\
      &+ \pR \cdot \vec{J}[\phi,-\vec{A}]
       + \frac{2 i}{\hbar} \vec{\Pi}[\chi] \cdot \vec{J}[\phi,-\vec{A}].
       \label{eq:cdce}
  \end{align}
  It was already introduced and discussed in \cite{schild2018}, and it is similar to its time-dependent counterpart \eqref{eq:tdce} because of the correspondences
  \begin{align}
    \frac{1}{M} \vec{P}[\chi,\vec{A}] \cdot \pR \rho(\vec{r}|\vec{R}) & \leftrightarrow \pt \rho(\vec{r}|t) \\
    \pr \cdot \vec{j}[\phi](\vec{r}|\vec{R}) & \leftrightarrow \pr \cdot \vec{j}(\vec{r}|t).
  \end{align}
  The terms including the system(clock) flux density $\vec{J}[\phi,-A]$ appear only in the clock-dependent treatment.
  This flux density reflects the fact that the state of the system depends on configuration of the clock, and there is also a flux associated along the different clock configurations that are distributed according to the probability density $|\chi|^2$ of the clock.
  
  The continuity equation \eqref{eq:cdce} was written such that it resembles 
  its time-dependent counterpart \eqref{eq:tdce} as closely as possible.
  However, a more natural way of writing \eqref{eq:cdce} is
  \begin{align}
    0 =& \frac{1}{M} \left( \vec{P}[\chi,\vec{A}] \cdot \vec{\Pi}[\phi]
                          + \vec{\Pi}[\chi] \cdot \vec{P}[\phi,-\vec{A}]
                \right) \nonumber \\
      &+ \frac{1}{m} \left( \vec{\pi}[\phi] \cdot \vec{p}[\phi] + \frac{\popc \cdot \vec{p}[\phi]         }{2} \right) \nonumber \\
      &+ \frac{1}{M} \left( \vec{\Pi}[\phi] \cdot \vec{P}[\phi,-\vec{A}] + \frac{\Popc \cdot \vec{P}[\phi,-\vec{A}]}{2} \right),
      \label{eq:cdce1}
  \end{align}
  which is entirely in terms of the momentum fields (compare to \eqref{eq:tdce1}
  for the time-dependent case).
  Writing the continuity equation as \eqref{eq:cdce1} allows to compare it 
  directly with the clock-dependent Hamilton-Jacobi equation \eqref{eq:cdhje}.
  It is apparent that in the latter, only products of two classical momentum 
  fields ($\vec{p}[\dots]$ or $\vec{P}[\dots]$) or two quantum momentum fields 
  ($\vec{\pi}[\dots]$ or $\vec{\Pi}[\dots]$) appear, as well as the second 
  derivatives of the quantum momentum fields.
  In contrast, in the clock-dependent continuity equation only mixed products 
  of one classical and one quantum momentum field are found, as well as the 
  second derivatives of the classical momentum fields.
  
  After having developed the theory of CDQHD, we can turn to its solution in terms of trajectories.
  The clock-dependent Hamilton Jacobi equation \eqref{eq:cdhje} can be interpreted as differential equation for the phase $\arg(\phi)$ alone.
  Given the momentum fields, it can be solved in terms of trajectories by solving 
  \begin{align}
    \frac{d}{d\tau} \vec{r}_{\rm t}(\tau)    
      =& \frac{\vec{p}_{\rm t}[\phi]}{m}   \label{eq:pt1}\\
    \frac{d}{d\tau} \vec{R}_{\rm t}(\tau)    
      =& \frac{\vec{P}_{\rm t}[\chi,A] + \vec{P}_{\rm t}[\phi,-A]}{M} 
      = \frac{\vec{P}_{\rm t}[\psi]}{M} \label{eq:pt2}
  \end{align}
  where $\tau$ is a parameter along the trajectory and the subscript ``${\rm t}$'' means that the corresponding quantity is evaluated along the trajectory.
%   Equations \eqref{eq:pt1} and \eqref{eq:pt2} are only the equations for the 
%   paths in $(\vec{r}, \vec{R})$-space, the full set of equations for the 
%   characteristics is given in the appendix.
  The configuration $\vec{r}_{\rm t}$ depends on $\tau$ and, conditionally, on the trajectory $\vec{R}_{\rm t}$ of the clock via the dependence of the momentum field $\vec{p}$ on the conditional wavefunction $\phi(r|\vec{R}_{\rm t})$.
  The coordinate of the clock, $\vec{R}_{\rm t}$, is determined by the momentum field of both the clock and the system(clock), reflecting the fact that clock and system are, in general, coupled.
%   , which corresponds to the (gauge-invariant) momentum field of the state $\psi$ of the  combined system w.r.t.\ the clock coordinates.
  
  So far, we treated the interaction of a quantum clock with a quantum system that, together, form a closed system described by the static Schr\"odinger equation \eqref{eq:tise_psi}
  However, the experimental situations of interest is different:
  To induce and to probe an electron dynamics in molecules, a controlled interaction with an external potential (e.g.\ a laser field) is necessary.
  This leads to an introduction of a time parameter, and the electron-nuclear system is described by a TDSE.
  The treatment of Sec.\ \ref{sec:cdse} can be generalized to include such an external time by starting from a TDSE instead of the Schr\"odinger equation \eqref{eq:tise_psi}, see e.g.\ \cite{abedi2010}.
  The equations for the clock-and-time-dependent quantum hydrodynamics that follow are then a combination those of Sec.\ \ref{sec:tdqhd} for TDQHD and those of Sec.\ \ref{sec:cdqhd} for CDQHD.
  However, for the trajectories it turns out that the parameter $\tau$ in \eqref{eq:pt1} and \eqref{eq:pt2} can simply be identified with the external time $t$.
  It thus suffices to solve these equations to obtain quantum trajectories for the dynamics of electrons in molecules.
  
  \section{Electron trajectories with quantum nuclei}
  \label{sec:example}
  
  Having establish how quantum hydrodynamics depends on a classical time and on a quantum clock, we can use this knowledge to obtain electronic quantum trajectories in molecules.
  The molecular wavefunction is $\psi(\vec{R},\vec{r}|t)$, and it depends on the nuclear coordinates $\vec{R}$, the electronic coordinates $\vec{r}$, and a classical time parameter $t$.
  The classical time parameter can originate e.g.\ from interaction with a laser field or other time-dependent potentials that constitute experimental control parameters.
  The electronic wavefunction $\phi(\vec{r}|\vec{R},t)$ depends conditionally on the nuclear configuration and both $\phi$ as well as the nuclear wavefunction is $\chi(\vec{R}|t)$ depend conditionally on $t$.
  Following the argument of Sec.\ \ref{sec:cdse}, the nuclei behave as quantum clock for the electrons.
  As the discussion of Sec.\ \ref{sec:cdqhd} shows, we only need to solve \eqref{eq:pt1} and \eqref{eq:pt2} to obtain the electronic trajectories.
  
%   To illustrate these trajectories, the model of \cite{eich2016} for a proton-coupled electron transfer is used that has also previously been studied in connection with the CDCE \cite{schild2018}.
%   It is a model for a dynamics of an electronic and a nuclear degree of freedom, where the dynamics itself is generated from a TDSE that refers to an external time parameter $t$.
%   Hence, the model is a model for time- and clock-dependent quantum hydrodynamics with the quantum system being the electronic degree of freedom, the quantum clock being the nuclear degree of freedom, and with time $t$ defined by an
%   unspecified but essentially classical clock, e.g.\ a laser field that initiates the dynamics and that might be used to also probe the dynamics.
  
  \begin{figure}[htbp]
    \centering
    \includegraphics[width=0.99\textwidth]{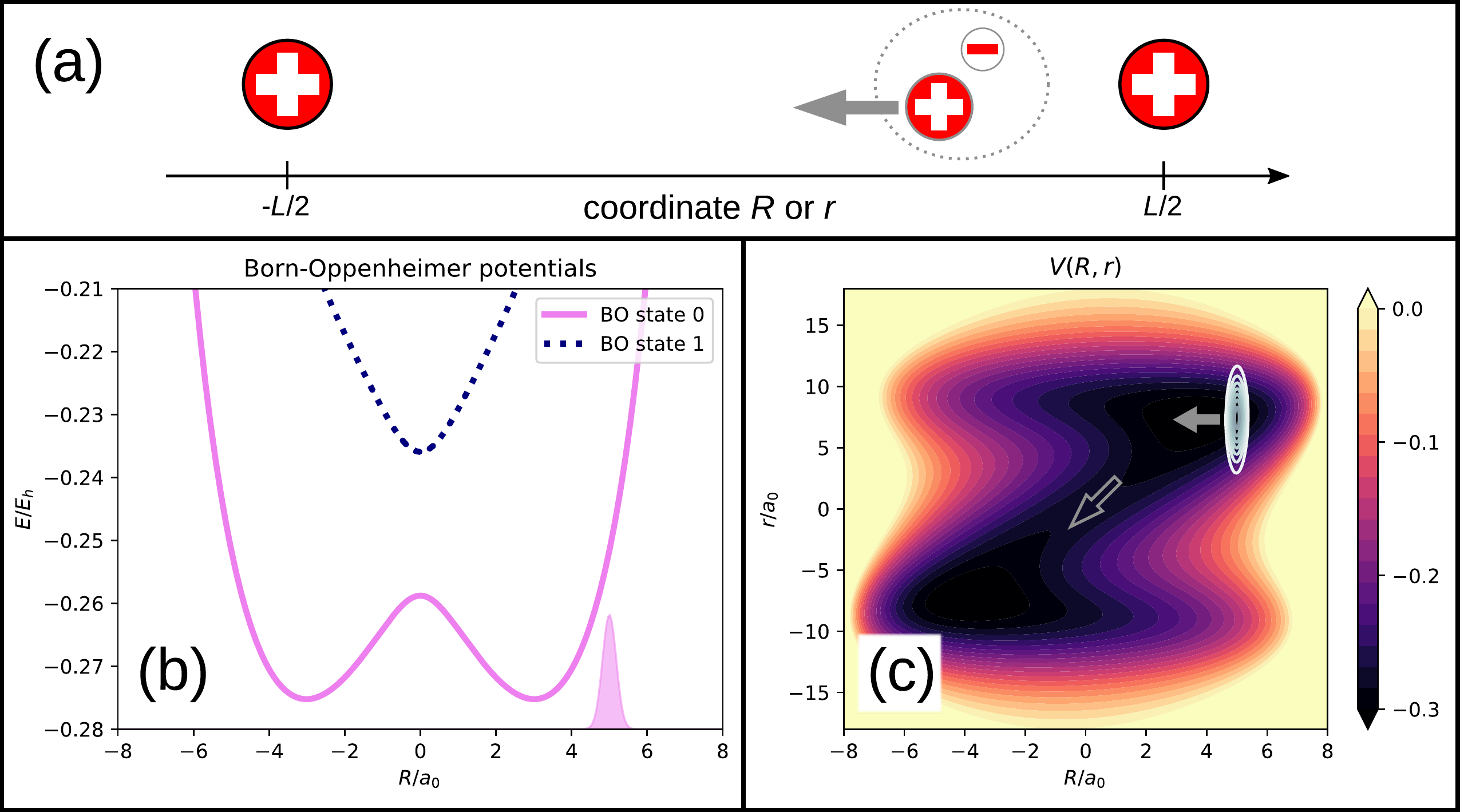}
    \caption{
      (a) A heavy particle with positive charge (nucleus) moves along $R$ and an electron moves along $r$, while two positive charges are clamped at $R = \pm L$, $r = \pm L$. 
      The dynamics is such that the nucleus starts close to a clamped positive charge, is repelled and moves toward smaller $R$ while dragging the electron with it (toward smaller $r$).
      (b) The lowest two Born-Oppenheimer potentials and initial marginal density of the  for the nucleus. 
      (c) Model potential $V(R,r)$ and initial wavefunction electron-nuclear wavefunction. 
      Two arrows indicate the motion during the dynamics.
      }
    \label{fig:potentials}
  \end{figure}
  
  To illustrate these trajectories, the model of \cite{eich2016} for a proton-coupled electron transfer is used that has also previously been studied in connection with the CDCE \cite{schild2018}.
  It is a model for a dynamics of one electronic and a nuclear degree of freedom sketched in Fig.\ \ref{fig:potentials}a.
  A heavy positively charged particle (the nucleus) moves along coordinate $R$, and a light negatively charged particle (the electron) moves along coordinate $r$.
  Two clamped (infinitely heavy) positive charges are located at 
  $(R,r) = (\pm L/2, \pm L/2)$.
  The model Hamiltonian is
  \begin{align}
    H = -\frac{\mu}{2} \partial_R^2 + \hat{H}_{\rm S} 
  \end{align}
  with $\mu = m/M$ being the mass ratio between the electron and the nucleus, where
  \begin{align}
    \hat{H}_{\rm S} = -\frac{\partial_r^2}{2} + V(R,r)
    \label{eq:hs}
  \end{align}
  contains the kinetic energy of the electron and the scalar interaction potential
  \begin{align}
    V(R,r) =  \frac{1}{|R - \frac{L}{2}|}
                              + \frac{1}{|R + \frac{L}{2}|}
                              - \frac{\operatorname{erf} \left( \frac{|r-R|}{R_{\rm c}} \right)}{|R - r|}
                              - \frac{\operatorname{erf} \left( \frac{|r-\frac{L}{2}|}{R_{\rm r}} \right)}{|r - \frac{L}{2}|}
                              - \frac{\operatorname{erf} \left( \frac{|r+\frac{L}{2}|}{R_{\rm l}} \right)}{|r + \frac{L}{2}|}.
                              \label{eq:vrr}
  \end{align}
  We use the parameters \unit[$L= 19$]{$a_0$}, \unit[$R_{\rm r} = R_{\rm l} = 3.5$]{$a_0$},
  \unit[$R_{\rm c} = 4.0$]{$a_0$}, and a mass ratio $\mu^{-1} = 900$.
  For these model parameters, the Born-Oppenheimer potential energy surfaces 
  $\epsilon_n^{\rm BO}(R)$ for the nucleus, given by
  \begin{align}
    \hat{H}_{\rm S} \phi_n^{\rm BO}(r|R) = \epsilon_n^{\rm BO}(R) \phi_n^{\rm BO}(r|R),
    \label{eq:hsboa}
  \end{align}
  are energetically well-separated, as shown in Fig.\ \ref{fig:potentials}b.
  The dynamics is adiabatic in the sense that 
  \begin{align}
    |\phi(r|R,t)|^2 \approx |\phi_0^{\rm BO}(r|R)|^2,
    \label{eq:adia}
  \end{align}
  i.e., the density of the electronic wavefunction does not depend on the external time $t$ but only on the nuclear configuration $R$.
  The phase of $\phi$ depends on $t$, however, and leads to electronic quantum trajectories.
  
  The initial state for the dynamics is chosen to be a product of a Gaussian 
  and the ground-state Born-Oppenheimer electronic wavefunction,
  \begin{align}
    \psi_0(R,r) = \chi_0(R) \phi_0^{\rm BO}(r|R),
  \end{align}
  where 
  \begin{align}
    \chi_0(R) \propto e^{-\frac{(R-R_0)^2}{4 \sigma^2}}
  \end{align}
  with \unit[$R_0 = 5$]{$a_0$} and \unit[$\sigma = 0.183$]{$a_0$}.
  The density $|\chi_0(R)|^2$ is also shown in Fig.\ \ref{fig:potentials}b.
  From the figure, it is clear that this wavepacket 
  will initially move toward smaller $R$ and has enough energy to overcome the 
  barrier at $R=0$.
  In Fig.\ \ref{fig:potentials}c, the density $|\psi_0(R,r)|^2$ for the initial state is shown together with the potential $V(R,r)$ of \eqref{eq:vrr}.
  The dynamics is indicated by two arrows.
  The nucleus moves from its initial center at \unit[$R = 5$]{$a_0$} toward smaller $R$ and the electron follows from its initial center at ca.\ \unit[$r = 8$]{$a_0$} toward smaller $r$, resulting in the motion indicated by the arrows.
  
  \begin{figure}[htbp]
    \centering
    \includegraphics[width=0.99\textwidth]{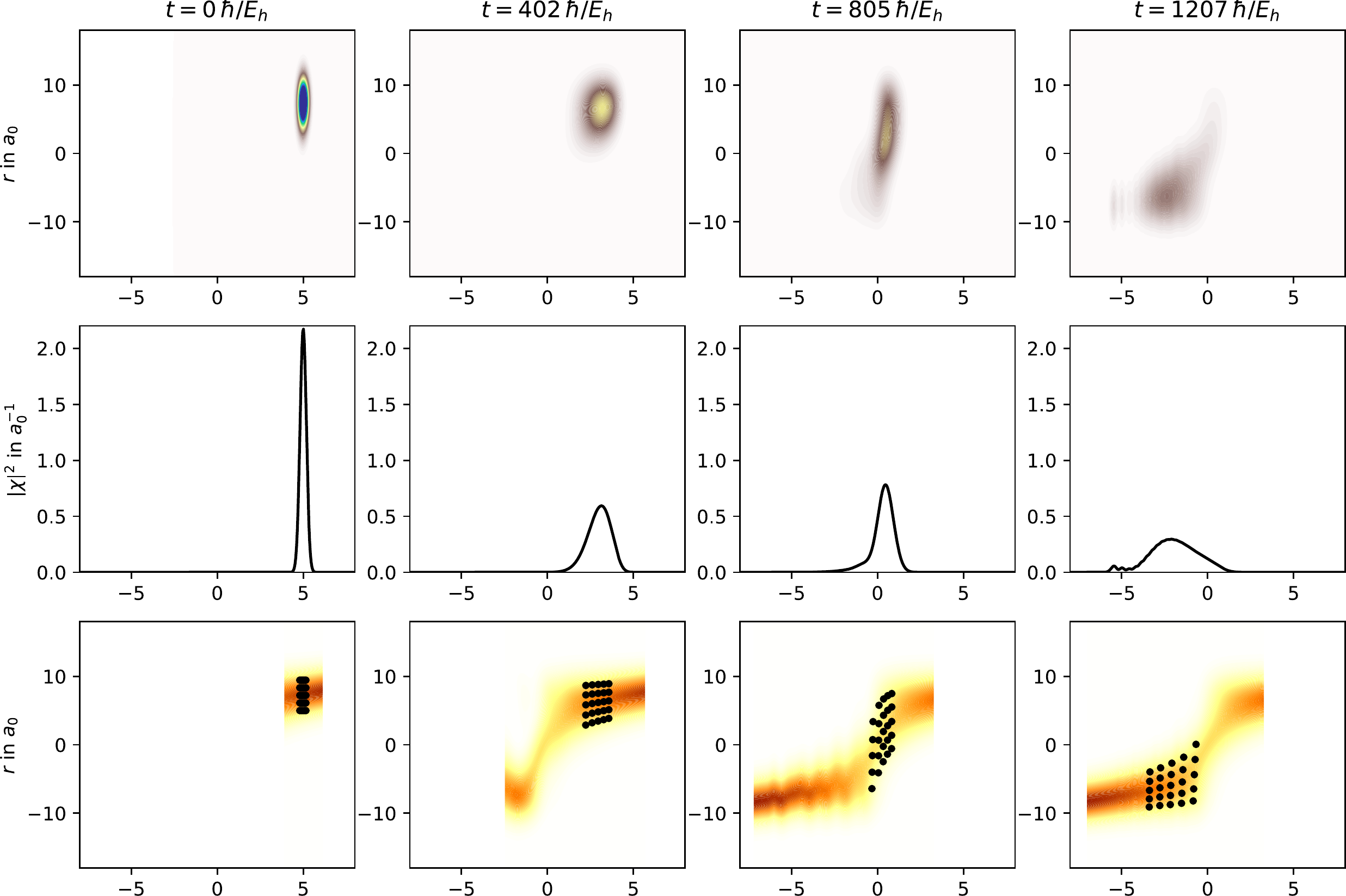}
    \caption{
      Density of the joint electron-nuclear wavefunction $\psi(R,r|t)$ (top), of the marginal nuclear wavefunction $\chi(R|t)$ (middle), and of the conditional electronic wavefunction $\phi(r|R,t)$ together with the electronic quantum trajectories (bottom) for four different values of the external time $t$.}
    \label{fig:trajectories}
  \end{figure}
  
  Fig.\ \ref{fig:trajectories} shows the dynamics in terms of the joint 
  probability density $|\psi(R,r|t)|^2$ (top), the marginal probability density 
  of the nucleus $|\chi(R|t)|^2$ (center), and the conditional probability density 
  of the electron $|\phi(r|R,t)|^2$ (bottom), for four different values of the 
  external time $t$.
  The joint probability density follows the motion indicated by the arrows in 
  Fig.\ \ref{fig:potentials}c, i.e., its maximum moves first toward smaller $R$,
  indicating the nuclear motion, but then also toward smaller $r$, indicating 
  the electron following the nucleus.
  The motion toward smaller $R$ is clearly visible in the marginal nuclear
  density $|\chi|^2$.
  
  The probability density for our actual system of interest, the electron, is the conditional density $|\phi|^2$ shown as contour plot in the bottom panels of Fig.\ \ref{fig:trajectories}.
  It is normalized for each value of $R$ (and $t$), hence it is localized somewhere along $r$ for every $R$.
  In the figure, $|\phi|^2$ is only shown in a region of $R$ where the probability density of finding the nucleus is \unit[$|\chi|^2 > 10^{-8}$]{a$_0^{-1}$}, as its computation via \eqref{eq:phidef} is numerically problematic for very small $|\chi|$.
  
  Electronic trajectories were computed for the wavefunction $\phi$ using \eqref{eq:pt1} and \eqref{eq:pt2}, where $\tau$ is identified with the external time $t$.
  The initial coordinates of the nucleus $R_{\rm t}(t=0)$ and the electron $r_{\rm t}(t=0,R_{\rm t}(t=0))$ are chosen to be in a region where the density of the joint system $|\psi(R,r|t=0)|^2$ is large.
  Then, \eqref{eq:pt2} is solved to propagate the coordinate of the nucleus $R_{\rm t}(t)$ along the trajectory.
  Thereafter, \eqref{eq:pt1} is solved to propagate the electron $r_{\rm t}(t,R_{\rm t}(t))$ to the next value of $t$ and of $R_{\rm t}$, thus providing trajectories of the electron for each trajectory of the nucleus.
  The trajectories closely follow $|\phi|^2$ with $t$, illustrating that these are indeed the conditional trajectories corresponding to the state of the electron, given the nucleus is at a certain coordinate, and given a value for the external time $t$.
  
  \section{Discussion}
  \label{sec:discussion}
  
  We have illustrated the computation of electronic quantum trajectories for quantum nuclei with a model where the full quantum dynamics can be solved numerically.
  In this case, the wavefunction is available and electronic trajectories can be used for analysis of the electron dynamics.
  Such an analysis is particularly interesting when a reaction mechanism is of interest and we want how the electronic rearrangement takes place during a chemical reaction.
  The reaction mechanism can then be obtained from the electronic flux density or from electron trajectories, where the latter allow for a straightforward interpretation of the dynamics.
  A quantum feature of the nuclei in this context can e.g.\ be a branching of the nuclear wavepacket into different parts that may correspond to different reaction mechanisms.
  This article presents the theory to describe such a situation.
  
  To compute electronic quantum trajectories for complex systems from a given electronic wavefunction, it is necessary to obtain the phase of this wavefunction.
  For a dynamics that is well described in the BOA, like our example dynamics, \eqref{eq:adia} holds and electronic phases are typically not computed; the electronic subsystem provides the potential energy surfaces for the nuclei but the electronic motion itself is neglected.
  A way to recover the electron dynamics approximately is nuclear velocity perturbation theory \cite{scherrer2013,schild2016}, where a simplified version of the CDSE is solved.
  Such a computation is feasible for ab initio molecular dynamics simulations of medium-sized molecules (see \cite{debus2018} for an example), hence electronic quantum trajectories can be calculated for such simulations and information about the electronic reaction mechanism can be obtained.
  Also, if the nuclear dynamics is obtained from multiple potential energy surfaces, the phase of the electronic wavefunction and, consequently, also electronic quantum trajectories are readily available (see e.g.\ \cite{pohl2017}).
  
  As a computational tool, trajectory calculations can, in principle, be more efficient than a wavepacket propagation.
  In practice, however, there are computational problems for quantum trajectories that are rooted in the existence of the quantum potential and the fact that the propagation of one trajectory is coupled to the other trajectories \cite{wyatt2005}.
  As discussed in the introduction, many approximation methods have been developed for nuclear trajectories but little has been done for electronic trajectories.
  One reason for this difference is the success of wavefunction- or density-based electronic structure calculations and the focus on nuclear dynamics.
  With the advent of ultrafast laser pulses and attoscience, it is  possible to directly measure the dynamics of the electronic time scale.
  Thus, interest is shifting to the dynamics of the electrons.
  A further development of a trajectory-based view on the quantum dynamics of electrons can be a step forward to overcome the current limitations of the simulation methods used in attoscience.
  
  In summary, electronic quantum trajectories are a useful tool for analysis and have the potential to also become computational tools.
  We presented the theory to account for nuclear quantum effects and we illustrated the computation of electronic trajectories.
  With this, we opened the way for further developments of a fully quantum-mechanical trajectory-based view on the dynamics of molecules and on the fundamentals of chemistry and molecular physics.
  
  {\bf Acknowledgment}
  
  This research is supported by an Ambizione grant of the Swiss National Science 
  Foundation.
  
  \bibliography{bib}{}
  \bibliographystyle{unsrt}

\end{document}